\documentclass{article}
\usepackage{graphicx}
\usepackage{amsmath}
\usepackage{amsfonts}
\usepackage{amssymb}

\begin{document}

\title{Superluminal Localized Waves of Electromagnetic Field in Vacuo}
\author{Peeter Saari\\Institute of Physics, University of Tartu,\\Riia 142, Tartu 51014, Estonia}
\maketitle
\begin{abstract}
Presented is an overview of electromagnetic versions of the so-called X-type
waves intensively studied since their invention in early 1990.-ies in
ultrasonics. These waves may be extremely localized both laterally and
longitudinally and -- what has been considered as most startling -- propagate
superluminally without apparent spread. Spotlighted are the issues of the
relativistic causality, variety of mathematical description and possibilities
of practical applications of the waves.
\end{abstract}

PACS numbers: 42.25.Bs, 03.40.Kf, 42.65.Re, 41.20.Jb.

\textbf{TO BE PUBLISHED IN} Proceedings of the conference ''Time's Arrows,
Quantum \ \ \ Measurements and Superluminal Behaviour'' (Naples, October 2-6,
2000) by the Italian NCR.

\section{Introduction}

More often than not some physical truths, as they gain general acceptance,
enter textbooks and become stock rules, loose their exact content for the
majority of the physics community. Moreover, in this way superficially
understood rules may turn to superfluous taboos inhibiting to study new
phenomena. For example, conviction that ''uniformly moving charge does not
radiate'' caused a considerable delay in discovering and understanding the
Cherenkov effect. By the way, even the refined statement ''uniformly moving
charge does not radiate in vacuum'' is not exact as it excludes the so-called
transition radiation known an half of century only, despite it is a purely
classical effect of macroscopic electrodynamics.

In this paper we give an overview of electromagnetic versions of the so-called
X-type waves intensively studied since 1990.-ies~\cite{Lu1X}-\cite{ItaalExp}.
The results obtained have encountered such taboo-fashioned attitudes
sometimes. Indeed, these waves, or more exactly -- wavepackets, may be
extremely localized both laterally and longitudinally and, what is most
startling, propagate superluminally without apparent diffraction or spread as
yet. Furthermore, they are solutions - although exotic - of linear wave
equations and, hence, have nothing to do with solitons or other localization
phenomena known in contemporary nonlinear science. Instead, study of these
solutions has in a sense reincarnated some almost forgotten ideas and findings
of mathematical physics of the previous turn of the century. X-type waves
belong to phenomena where a naive superluminality taboo ''group velocity
cannot exceed the speed of light in vacuum'' is broken. In this respect they
fall into the same category as plane waves in dispersive resonant media and
the evanescent waves, propagation of which (photon tunneling) has provoked
much interest since publication of papers~ \cite{Tunnel1},\cite{Tunnel2}%
,\cite{Tunnel3} .Therefore it is not surprising that tunneling of X waves in
frustrated internal reflection has been treated in a recent theoretical
paper~\cite{XjaFIR} .

Indeed, studies conducted in different subfields of physics, which are dealing
with superluminal movements, are interfering and merging fruitfully. A
convincing proof of this trend is the given Conference and the collection of
its papers in hand.

This is why in this paper we spotlight just superluminality of the X waves,
which is now an experimentally verified fact~\cite{OptLettmeie2}%
,\cite{PRLmeie},\cite{ItaalExp}, but which should not be considered as their
most interesting attribute in general. Their name was coined within
theoretical ultrasonics by the authors of the paper \cite{Lu1X} which
initiated an intensive study of the X waves, particularly due to outlooks of
application in medical ultrasonic imaging. Possible superluminality of
electromagnetic localized waves was touched by the authors of
Ref.~\cite{ZioMAIN} -- who had derived the waves under name ''slingshot
pulses'' independently from the paper~\cite{Lu1X} -- and became the focus of
growing interest thanks to E .Recami (see~\cite{RecamiMain} and references
therein), who pointed out physically deeply meaningful resemblance between the
shape of the X waves and that of the tachyon~\cite{RecamiTachkuju}. The
paper~\cite{RecamiTachkuju} was published in times of great activity in
theoretical study of these hypothetical superluminal particles. To these years
belongs paper~\cite{EMtach} where a double-cone-shaped ''electromagnetic
tachyon'' as a result of light reflection by a conical mirror was considered.
This a quarter-of-century-old paper seems to be the very pioneering work on
X-waves, though this and the subsequent papers of the same author have been
practically unknown and only very recently were rediscovered for the X wave
community (see references in the review~\cite{Brasiil}). Last but not least,
if one asked what was the very first sort of superluminal waves implemented in
physics, the answer would be -- realistic plane waves. Indeed, as it is well
known, the most simple physically feasible realization of a plane wave beam is
the Gaussian beam with its bounded cross-section and, correspondingly, a
finite energy flux. However, much less is known that due to the Gouy phase
shift the group velocity in the waist region of the Gaussian beam is slightly
superluminal, what one can readily check on the analytical expressions for the
beam (see also Ref.~\cite{Gouy}). For all the reasons mentioned, in this paper
we present -- after an introduction of the physical nature of the X-type waves
(Section 2) - quite in detail a new representation of the localized waves
(Section 3). This representation -- what we believe is a new and useful
addition into the theory of X-type waves -- in a sense generalizes the Huygens
principle into superluminal domain and directly relies on superluminality of
focal behavior of any type of free-space waves, which manifests itself in the
Gouy phase shift. The startling superluminality issues are briefly discussed
in the last Section.

Figures showing 3-dimensional plots have been included for a vivid
comprehension of the spatio-temporal shape of the waves, however, only few of
the animations showed in the oral presentation had sense to be reproduced here
in the static black-and-white form. The bibliography is far from being
complete, but hopefully a number of related references can be found in other
papers of the issue in hand.

\section{Physical nature of X-type waves}

In order to make the physical nature of the X-type superluminal localized
waves better comprehensible, we first discuss a simple representation of them
as a result of interference between plane wave pulses.%

\begin{center}
\includegraphics[
trim=0.681782in 7.235637in 1.600384in 0.422792in,
height=6.0231cm,
width=12.0441cm
]%
{figure1s.ps}%
\end{center}

\begin{description}
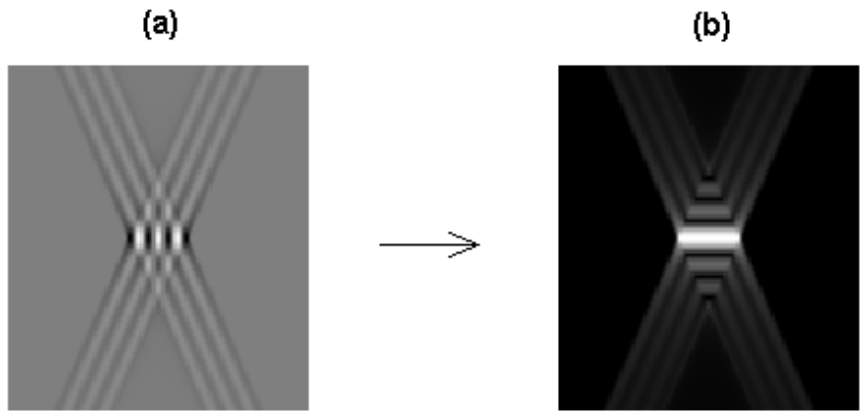
\item  Fig.1. X-type scalar wave formed by scalar plane wave pulses containing
three cosinusoidal cycles. The propagation direction (along the axis $z$ ) is
indicated by arrow. As linear gray-scale plots in a plane of the propagation
axis and at a fixed instant, shown are (a) the field of the wave (real part if
the plane waves are given as analytic signals) and (b) its amplitude
(modulus). Note that the central bullet-like part of the wave would stand out
even more sharply against the sidelobes if one plotted the the distribution of
the intensity (modulus squared) of the wave.
\end{description}

With reference to Fig.1(a) let us consider a pair of plane wave bursts
possessing identical temporal dependences and the wave vectors in the plane
$y=0$. Their propagation directions given by unit vectors $\mathbf{n}%
_{/}\mathbf{=}\left[  \sin\theta,\ 0,\ \cos\theta\right]  $ and $\mathbf{n}%
_{\backslash}\mathbf{=}\left[  -\sin\theta,\ 0,\ \cos\theta\right]  $ are
tilted under angle $\theta$ with respect to the axis $z.$ In spatio-temporal
regions where the pulses do not overlap their field is given simply by the
burst profile as $\Psi_{P}(\eta-ct)$, where $\eta$ is the spatial coordinate
along the direction $\mathbf{n}_{/}$ or $\mathbf{n}_{\backslash}$,
respectively. In the overlap region , if we introduce the radius vector of a
field point $\mathbf{r}=\left[  x,\ y,\ z\right]  $ the field is given by
superposition
\begin{align}
\Psi_{P}(\mathbf{r\,n}_{/}-ct)+\Psi_{P}(\mathbf{r\,n}_{\backslash}-ct)  &
=\Psi_{P}(x\sin\theta+z\cos\theta-ct)+\nonumber\\
&  +\Psi_{P}(-x\sin\theta+z\cos\theta-ct)\ , \label{2tasalainet}%
\end{align}
\ which is nothing but the well-known two-wave-interference pattern with
doubled amplitudes. Altogether, the superposition of the pulse pair -- as two
branches
$\backslash$%
and / form the letter X -- makes up\ an X-shaped propagation-invariant
interference pattern moving along the axis $z$ with speed $v=c/\cos\theta$
which is both the phase and the group velocity of the wave field in the
direction of the propagation axis $z$. This speed is superluminal in a similar
way as one gets a faster-than-light movement of a bright stripe on a screen
when a plane wave light pulse is falling at the angle $\theta$ onto the screen
plane. Let us stress that here we need not to deal with the vagueness of the
physical meaning inherent to the group velocity in general -- simply the whole
spatial distribution of the field moves rigidly with $v$ because the time
enters into the Eq.(\ref{2tasalainet}) only together with the coordinate $z$
through the propagation variable $z_{t}=z-vt$.

Further let us superimpose axisymmetrically all such pairs of waves whose
propagation directions form a cone around the axis $z$ with the top angle
$2\theta$, in other words, let the pair of the unit vectors be $\mathbf{n}%
_{/}\mathbf{=}\left[  \sin\theta\cos\phi,\ \sin\theta\sin\phi,\ \cos
\theta\right]  $ and $\mathbf{n}_{\backslash}\mathbf{=}\left[  \sin\theta
\cos\left(  \phi+\pi\right)  ,\ \sin\theta\sin\left(  \phi+\pi\right)
,\ \cos\theta\right]  $, where the angle $\phi$ runs from $0$ to $180$
degrees. As a result, we get an X-type supeluminal localized wave in the
following simple representation%

\begin{equation}
\Psi_{X}\left(  \rho,z_{t}\right)  =\int_{0}^{\pi}d\phi\ \left[  \Psi
_{P}\left(  \mathbf{r}_{t}\mathbf{n}_{/}\right)  +\Psi_{P}\left(
\mathbf{r}_{t}\mathbf{n}_{\backslash}\right)  \right]  =\int_{0}^{2\pi}%
d\phi\ \Psi_{P}\left(  \mathbf{r}_{t}\mathbf{n}_{/}\right)  \ ,
\label{BXfieldfin}%
\end{equation}
where $\mathbf{r}_{t}=\left[  \rho\cos\varphi,\ \rho\sin\varphi,\ z_{t}%
\right]  $ is the radius vector of a field point in the copropagating frame
and cylindrical coordinates $(\rho,\varphi,z)$ have been introduced for we
restrict ourselves in this paper to axisymmetric or so-called zeroth-order
X-type waves only. Hence, according to Eq.(\ref{BXfieldfin}) the field is
built up from interfering pairs of identical bursts of plane waves. Fig.1
gives an example in which the plane wave profile $\Psi_{P}$ contains three cycles.

The less extended the profile $\Psi_{P}$, the better the separation and
resolution of the branches of the X-shaped field. In the superposition the
points of completely constructive interference lie on the $z$ axis, where the
highly localized energy ''bullet'' arises in the center, while the intensity
falls off as $\rho^{-1}$ along the branches and much faster in all other
directions (note that in contrast with Fig.1(a) in the case of interference of
only two waves the on-axis and off-axis maxima must be of equal strength). The
optical carrier manifests itself as one or more (depending on the number of
cycles in the pulse) halo toroids which are nothing but residues of the
concentric cylinders of intensity characteristic of the Bessel beam. That is
why we use the term 'Bessel-X pulse' (or wave) to draw a distinction from
carrierless X waves. By making use of an integral representation of the
zeroth-order Bessel function $J_{0}\left(  v\right)  =\allowbreak\pi^{-1}%
\int_{0}^{\pi}\cos\left[  v\cos\left(  \phi-\varphi\right)  \right]  d\phi\ $,
where $\varphi$ is an arbitrary angle, and by reversing the mathematical
procedure described in Ref.~\cite{PRLmeie}, we get the common representation
of X-type waves as superposition of monochromatic cylindrical modes (Bessel
beams) of different wavenumber $k=\omega/c$%

\begin{equation}
\Psi_{X}(\rho,z,t)=\int_{0}^{\infty}dk\ S(k)\ J_{0}\left(  \rho k\sin
\theta\right)  \ \exp\left[  i\left(  zk\cos\theta-\omega t\right)  \right]
\ , \label{BX_k-esituses}%
\end{equation}
where $S(k)$ denotes the Fourier spectrum of the profile $\Psi_{P}$. Again,
the Eq.(\ref{BX_k-esituses}) gives for both the phase and the group velocities
(along the axis $z$ -- in the direction of the propagation of the packet of
the cylindrical waves) the superluminal value $v=c/\cos\theta$.

\section{X-waves as wakewaves}

Although the representation Eq.(\ref{BXfieldfin}) of the X-type waves as built
up from two-plane-wave-pulse interference patterns constitutes an easily
comprehensible approach to the superluminality issues, it may turn out to be
counter-intuitive for symmetry considerations as will be shown below. In this
section we develope another representation introduced in Ref.~ \cite{UPGarm},
which is, in a sense, a generalization of the Huygens principle into
superluminal domain and allows figuratively to describe formation of
superluminal localized waves.

As known in electrodynamics, the D'Alambert (source-free) wave equation
possesses a particular solution $D_{0}$, which is spherically symmetric and
can be expressed through retarded and advanced Green functions of the
equation, $G^{(+)}$ and $G^{(-)}$, respectively, as%

\begin{equation}
D_{0}=c^{-2}\left[  G^{(+)}-G^{(-)}\right]  =\frac{1}{4\pi Rc}\left[
\delta(R-ct)-\delta(R+ct)\right]  \ , \label{D0defn}%
\end{equation}
where $R$ is the distance from the origin and $\delta$ is the Dirac delta
function. Thus, the function $D_{0}$ represents a spherical delta-pulse-shaped
wave, first (at negative times $t$) converging to the origin (the right term)
and then (at positive times $t$) diverging from it. The minus sign between the
two terms, which results from the requirement that a source-free field cannot
have a singular point $R=+0$, is of crucial importance as it assures vanishing
of the function at $t=0$. This change of the sign when the wave goes through
the collapsed stage at the focus is also responsible for the $90$ degrees
phase factor associated with the Huygens-Fresnel-Kirchhoff principle and for
the Gouy phase shift peculiar to all focused waves. From Eq.(\ref{D0defn}),
using the common procedure one can calculate Li\'{e}nard-Wiechert\ potentials
for a moving point charge $q$ flying, e.g. with a constant velocity $v$ along
axis $z$. However, as $D_{0}$\ includes not only the retarded Green function
but also the advanced one and therefore what is moving has to be considered as
a source coupled with a sink at the same point. For such a Huygens-type source
there is no restriction $v\leq c$ and for superluminal velocity $v/c>1$ we
obtain axisymmetric scalar and vector potentials (in CGS units, Lorentz
gauge)$\ \Phi(\rho,z,t)$ and $\mathbf{A}(\rho,z,t)=\Phi(\rho,z,t)\mathbf{v}%
/c$, where $\rho$ is the radial distance of the field point from the axis $z$,
the velocity vector is $\mathbf{v=}[0,0,v]$, and%

\begin{equation}
\Phi(\rho,z,t)=\frac{2q}{\sqrt{(z-vt)^{2}+\rho^{2}(1-v^{2}/c^{2})}}\left\{
\begin{array}
[c]{c}%
\Theta\left[  -(z-vt)-\rho\sqrt{v^{2}/c^{2}-1}\right]  -\\
-\Theta\left[  (z-vt)-\rho\sqrt{v^{2}/c^{2}-1}\right]
\end{array}
\right\}  \ , \label{SommerFi}%
\end{equation}
where $\Theta(x)$ denotes the Heaviside step function. Here for the sake of
simplicity we do not calculate the electromagnetic field vectors $\mathbf{E}$
and $\mathbf{H}$ (or\textbf{\ }$\mathbf{B}$\textbf{)}, neither will we
consider dipole sources and sinks required for obtaining non-axisymmetric
fields. We will restrict ourselves to scalar fields obtained as superpositions
of the potential given by the Eq.(\ref{SommerFi}). The first of the two terms
in the Eq.(\ref{SommerFi}) gives an electromagnetic Mach cone of the
superluminally flying charge $q$. In other words -- it represents nothing but
a shock wave emitted by a superluminal electron in vacuum, mathematical
expression for which was found by Sommerfeld three decades earlier than Tamm
and Frank worked out the theory of the Cherenkov effect, but which was
forgotten as an unphysical result after the special theory of relativity
appeared~\cite{GinzReview}. The second term in the Eq.(\ref{SommerFi})
describes a leading and reversed Mach cone collapsing into the superluminal
sink coupled with the source and thus feeding the latter. Hence, the
particular solution to the wave equation which is given by the
Eq.(\ref{SommerFi}) represents a double-cone-shaped pulse propagating rigidly
and superluminally along the axis $z$. In other words -- it represents an
X-type wave as put together from (i) the cone of incoming waves collapsing
into the sink, thereby generating a superluminal Huygens-type point source and
from (ii) wakewave-type radiation cone of the source. Let it be recalled that
the field given by the Eq.(\ref{SommerFi}) had been found for $\delta$-like
spatial distribution of the charge. That is why the field diverges on the
surface of the double cone or on any of its X-shaped generatrices given by
$(z-vt)=\pm~\rho\sqrt{v^{2}/c^{2}-1}$ and the field can be considered as an
elementary one constituting a base for constructing various X-type waves
through appropriate linear superpositions. Hence, any axisymmetric X-type wave
could be correlated to its specific\ (continuous and
time-dependent)\ distribution $\rho=\delta(x)\delta(y)\lambda(z,t)$\ of the
''charge'' (or the sink-and-source) with linear\ density $\lambda(z,t)$\ on
the propagation axis, while the superluminal speed of the wave corresponds to
the velocity $v$ of propagation of that distribution along the axis, i.e.
$\lambda(z,t)=\lambda(z-vt)$.

Let us introduce a superluminal version of the Lorentz transformation
coefficient $\gamma=1/\sqrt{v^{2}/c^{2}-1}=\cot\theta$, where $\theta$ is the
Axicon angle considered earlier and let us first choose the ''charge''
distribution $\lambda(z-vt)=\lambda(z_{t})$ as a Lorentzian. In this case the
field potential is given as convolution of Eq.(\ref{SommerFi}) with the
normalized distribution, which can be evaluated using Fourier and Laplace
transform tables:%

\begin{equation}
\Phi(\rho,z_{t})\otimes\frac{1}{\pi}\frac{\Delta}{z_{t}^{2}+\Delta^{2}%
}=-q\sqrt{\frac{2}{\pi}}\operatorname{Im}\left(  \frac{1}{\sqrt{(\Delta
-iz_{t})^{2}+\rho^{2}/\gamma^{2}}}\right)  \ , \label{FieldIm}%
\end{equation}
where $\Delta$ is the HWHM of the distribution and $z_{t}=z-vt$ is, as in the
preceding section, the axial variable in the co-propagating frame. The
resulting potential shown in Fig.2 (a) moves rigidly along the axis $z$ (from
left to right in Fig.2) with the same superluminal speed $v>c$. The plot (a)
depicts qualitatively also the elementary potential as far as the divergences
of the Eq.(\ref{SommerFi}) are smoothed out in the Eq.(\ref{FieldIm}).

\bigskip%

\begin{center}
\includegraphics[
trim=0.411206in 3.794837in 0.884675in 0.914762in,
height=12.0441cm,
width=12.0441cm
]%
{figure2s.ps}%
\end{center}

\begin{description}
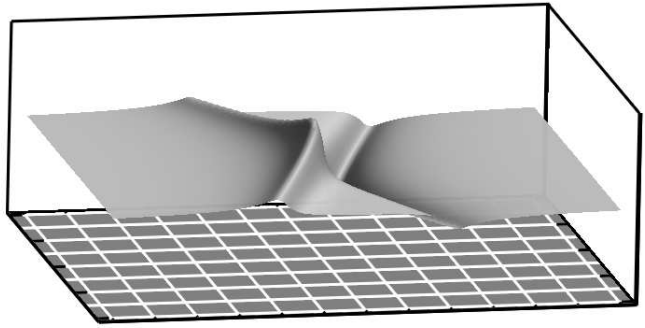
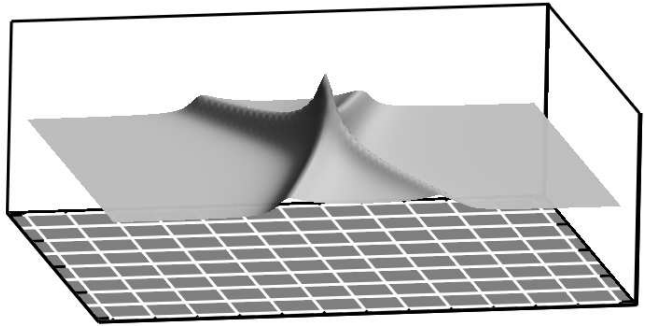
\item  Fig.2. Dependence on the longitudinal coordinate $z_{t\text{ }}=z-vt$
(increasing from the left to the right) and a lateral one $x=\pm\rho$ of the
imaginary (a) and real (b) parts of the field of the simplest X-wave. The
velocity $v=1.005c$ \ and, correspondingly, the superluminality parameter
$\gamma=10$. Distance between grid lines on the basal plane is $4\Delta$ along
the axis $z_{t\text{ }}$ and $20\Delta\ $along$\ $the lateral axis, the unit
being the half-width $\Delta$.
\end{description}

We see that an unipolar and even ''charge'' distribution gives an odd and
bipolar potential, as expected, while the symmetry of the plot differs from
what might be expected from superimposing two plane wave pulses under the tilt
angle $2\theta$. Indeed, in the latter case the plane waves are depicted by
each of the two diagonal branches (%
$\backslash$%
and /) of the X-shaped plot and therefore the profile of the potential on a
given branch has to retain its sign and shape if one moves from one side of
the central interference region to another side along the same branch.
Disappearance of the latter kind of symmetry, which can be most distinctly
followed in the case of bipolar single-cycle pulses -- just the case of Fig.2
a -- is due to mutual interference of all the plane wave pairs forming the
cone as $\phi$ runs from $0$ to $\pi$ .

Secondly, let us take the ''charge'' distribution as a dispersion curve with
the same width parameter $\Delta$ , i.e. as the Hilbert transform of the
Lorentzian. Again, using Fourier and Laplace transform tables, we readily obtain:%

\begin{equation}
\Phi(\rho,z_{t})\otimes\frac{1}{\pi}\frac{z_{t}}{z_{t}^{2}+\Delta^{2}}%
=q\sqrt{\frac{2}{\pi}}\operatorname{Re}\left(  \frac{1}{\sqrt{(\Delta
-iz_{t})^{2}+\rho^{2}/\gamma^{2}}}\right)  \ . \label{FieldRe}%
\end{equation}

The potential of the Eq.(\ref{FieldRe}) depicted in Fig.2 (b) is -- with
accuracy of a real constant multiplier -- nothing but the well-known
zeroth-order unipolar X wave, first introduced in Ref.~\cite{Lu1X} and studied
in a number of papers afterwards. Hence, we have demonstrated here how the
real and imaginary part of the simplest X-wave solution
\begin{equation}
\Phi_{X_{0}}(\rho,z_{t})\varpropto\frac{1}{\sqrt{(\Delta-iz_{t})^{2}+\rho
^{2}/\gamma^{2}}} \label{Xwave}%
\end{equation}
\ of the free-space wave equation can be represented as fields generated by
corresponding ''sink-and-source charge'' distributions moving superluminally
along the propagation axis. The procedure how to find for a given axisymmetric
X-type wave its ''generator charge distribution'' is readily derived from a
closer inspection of the Eq.(\ref{SommerFi}). Namely, on the axis $z$, i.~e.
for $\rho=0$, the Eq.(\ref{SommerFi}) constitutes the Hilbert transform kernel
for the convolution. Therefore, the ''charge'' distribution can be readily
found as the Hilbert image of the on-axis profile of the potential and vice versa.

Hence, we have obtained a figurative representation in which the superluminal
waves can be classified \textit{via} the distribution and other properties of
the Huygens-type sources propagating superluminally along the axis and thus
generating the wave field~\cite{slingshot}. Such representation -- which may
be named as Sommerfeld representation to acknowledge his unfortunate result of
1904 -- has been generalized to nonaxisymmetric and vector fields and applied
by us to various known localized waves \cite{Saari2bePub}.%

\begin{center}
\includegraphics[
trim=0.513961in 3.846123in 1.090371in 1.171737in,
height=12.0441cm,
width=12.0441cm
]%
{figure3s.ps}%
\end{center}

\begin{description}
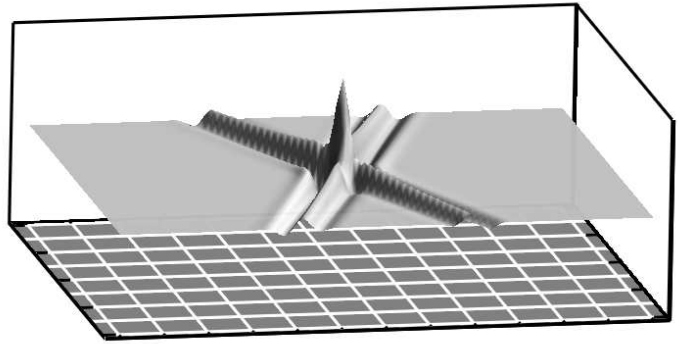
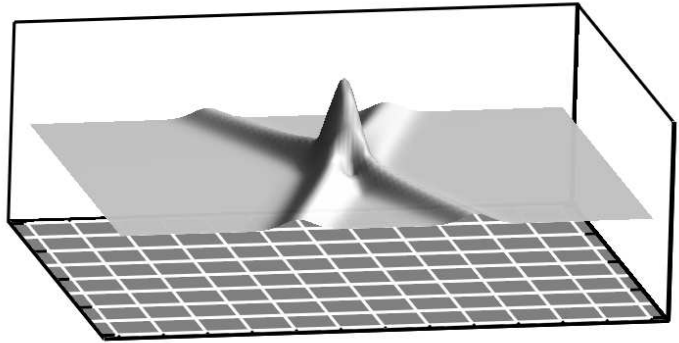
\item  Fig.3. The longitudinal-lateral dependences of the real part (a) and
the modulus (b) of the field of the Bessel-X wave. The parameters $v$ and
$\gamma$ are the same as in Fig.2. The new parameter of the wave -- the
wavelength $\lambda=2\pi/k_{z}$ of the optical carrier being the unit, the
distance between grid lines on the basal plane is $1\lambda$ along the axis
$z_{t\text{ }}$ and $5\lambda\ $along$\ $the lateral axis, while the
half-width $\Delta=\lambda/2$. For visible light pulses $\lambda$ is in
sub-micrometer range, which means that the period of the cycle as well as full
duration of the pulse on the propagation axis are as short as a couple of femtoseconds.
\end{description}

For example, in optical domain one has to deal with the so-called Bessel-X
wave~ \cite{Saari1}-\cite{PRLmeie}, which is a band-limited and oscillatory
version of the X wave. It is obvious that for the Bessel-X wave the ''charge''
distribution contains oscillations corresponding to the optical carrier of the
pulse. Bandwidth (FWHM) equal to ( or narrower than) the carrier frequency
roughly corresponds to 2-3 ( or more) distinguishable oscillation cycles of
the field as well as of the ''charge'' along the propagation axis.
Fortunately, few-cycle light pulses are affordable in contemporary femtosecond
laser optics. On the other hand, if the number of the oscillations in the
Bessel-X wave pulse (on the axis $z$ ) is of the order $n\simeq10$, the
X-branching occurs too far from the propagation axis, i.e. in the outer region
where the field practically vanishes and, with further increase of $n$, the
field becomes just a truncated Bessel beam. The analytic expression for a
Bessel-X wave depends on specific choice of the oscillatory function or,
equivalently, of the Fourier spectrum of the pulse on the axis $z$. One way to
obtain a Bessel-X wave possessing approximately $n$ oscillations is to take a
derivative of the order $m=n^{2}$ from the Eq.(\ref{Xwave}) with respect to
$z_{t}$ (or $z$ or $t$), which according to Eqs.(\ref{FieldIm}),(\ref{FieldRe}%
) is equivalent to taking the same derivative from the distribution function.
The $m$th temporal derivative of the common X wave can be expressed in closed
form through the associated Legendre polynoms~ \cite{FinLegendre}. Another way
is to use the following expression, which for $n\gtrsim3$ approximates well
the field of the Bessel-X wave with a near-Gaussian spectrum~\cite{Saari1},\cite{LaserPhys}

\bigskip%
\begin{equation}
\Phi_{BX_{0}}(\rho,z_{t})\varpropto\sqrt{Z(z_{t})}\exp\left[  -\frac{1}%
{\Delta^{2}}\left(  z_{t}^{2}+\rho^{2}/\gamma^{2}\right)  \right]  \cdot
J_{0}\left[  Z(z_{t})\,k_{z}\rho/\gamma\right]  \cdot\exp(ik_{z}z_{t})\ ,
\label{BX}%
\end{equation}
where complex-valued function $Z(z_{t})=1+i\cdot z_{t}/k_{z}\Delta^{2}$ makes
the argument of the Bessel function also complex .The longitudinal wavenumber
$k_{z}=k\cos\theta=(\omega/c)\cos\theta$\ together with the half-width
$\Delta$ (at $1/e$-amplitude level on the axis $z$) are the parameters of the
pulse. Again, dependence on $z,t$ through the single propagation variable
$\ z_{t}=z-vt$ indicates the propagation-invariance of the wave field shown in Fig.3.

\bigskip

\section{Application prospects of the X-type waves}

Limited aperture of practically realizable X-type waves causes an abrupt decay
of the interference structure of the wave after flying rigidly over a certain
distance. However, the depth of invariant propagation of the central spot of
the wave can be made substantial -- by the factor $\cot\theta=\gamma
=1/\sqrt{v^{2}/c^{2}-1}$ larger than the aperture diameter. Such type of
electromagnetic pulses, enabling directed, laterally and temporally
concentrated and nonspreading propagation of wavepacket energy through
space-time have a number of potential applications in various areas of science
and technology. Let us briefly consider some results obtained along this line.

Any ultrashort laser pulse propagating in a dispersive medium -- even in air
-- suffers from a temporal spread, which is a well-known obstacle in
femtosecond optics. For the Bessel-X wave with its composite nature, however,
there exists a possibility to suppress the broadening caused by the
group-velocity dispersion~\cite{Saari1},\cite{OptLettmeie1}. Namely, the
dispersion of the angle $\theta$, which is to a certain extent inherent in any
Bessel-X wave generator, can be played against the dispersion of the medium
with the aim of their mutual compensation. The idea has been verified in an
experimental setup with the lateral dimension and the width of the temporal
autocorrelation function of the Bessel-X wave pulses, respectively, of the
order of 20 microns and 200 fs~\cite{OptLettmeie2}. Thus, an application of
optical X-type waves has been worked out -- a method of designing femtosecond
pulsed light fields that maintain their strong (sub-millimeter range)
longitudinal and lateral localization in the course of superluminal
propagation into a considerable depth of a given dispersive medium.

Optical Bessel-X waves allow to accomplish a sort of diffraction-free
transmission of arbitrary 2-dimensional images~\cite{Saari1},\cite{LaserPhys}.
Despite its highly localized ''diffraction-free'' bright central spot, the
zeroth-order monochromatic Bessel beam behaves poorly in a role of
point-spread function in 2-D imaging. The reason is that its intensity decays
too slowly with lateral distance, i.e. as $\sim\rho^{-1}$. On the contrary,
the Bessel-X wave is offering a loop hole to overcome the problem. Despite the
time-averaged intensity of the Bessel-X wave possesses the same slow radial
decay $\sim\rho^{-1}$ due to the asymptotic behavior along the X-branches, an
instantaneous intensity has the strong Gaussian localization in lateral
cross-section at the maximum of the pulse and therefore it might serve as a
point-spread function with well-constrained support but also with an
extraordinary capability to maintain the image focused without any spread over
large propagation depths. By developing further this approach it is possible
to build a specific communication system~\cite{LuSideliin}. Ideas of using the
waves in high-energy physics for particle acceleration -- one of such was
proposed already two decades ago~\cite{Kolb}\ -- are not much developed as yet.

It is obvious that for a majority of possible applications the spread-free
central spot is the most attractive peculiarity of the X-type waves. The
better the faster the intensity decay along lateral directions and X-branches
is. In this respect a new type of X waves -- recently discovered Focused X
wave~\cite{PIEReview} -- seems to be rather promising.

As one can see in Fig.4. and by inspecting an analytical expression for the wave%

\begin{equation}
\Phi_{FX_{0}}(\rho,z,t)=\frac{\Delta\exp(k_{0}\gamma\Delta)}{R(\rho,z_{t}%
)}\exp\left[  ik_{0}\gamma\left[  iR(\rho,z_{t})+(v/c)z-ct\right]  \right]
\;, \label{FX}%
\end{equation}

where $R(\rho,z_{t})=\sqrt{\left[  \Delta+iz_{t}\right]  ^{2}+\left(
\rho\,/\gamma\right)  ^{2}}$ and $k_{0}$ is a parameter of carrier wavenumber
type, the wave is very well localized indeed. However, like luminal localized
waves called Focus Wave Modes~\cite{ZioMAIN},\cite{PIEReview}, for this wave
propagation-invariant is the intensity only, while the wave field itself
changes during propagation in an oscillatory manner due to the last phase
factor $\exp\left[  ik_{0}\gamma\left[  (v/c)z-ct\right]  \right]  $ in the
Eq.(\ref{FX}), which has another $z,t$-dependence than through the propagation
variable $z_{t}=z-vt$. An animated version of Fig.4(a), made for the oral
presentation of the paper, shows that the oscillatory modulation moves in the
direction $-z$, i.e. opposite to the pulse propagation.%

\begin{center}
\includegraphics[
trim=0.565428in 3.923326in 1.193126in 1.248760in,
height=12.0441cm,
width=12.0441cm
]%
{figure4s.ps}%
\end{center}

\begin{description}
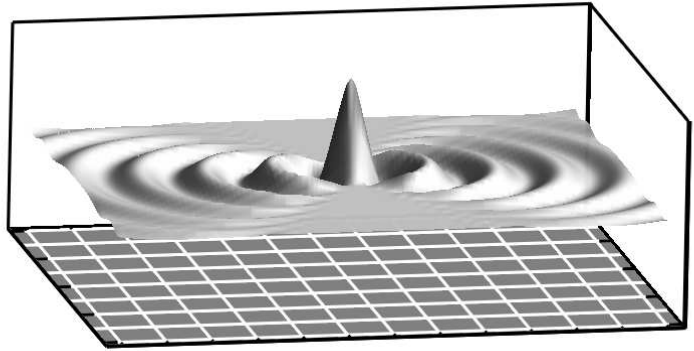
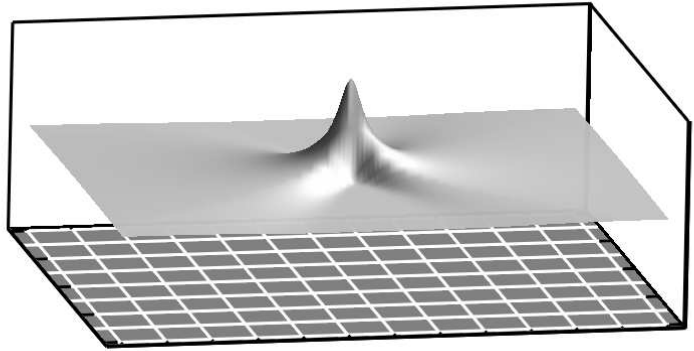
\item  Fig.4. The longitudinal-lateral dependences of the real part (a) and
the modulus (b) of the field of the Focused X wave. The parameters $v=1.01c$
and $\gamma=7$. Distance between grid lines on the basal plane is $4\Delta$
along both axes. $\lambda=2\pi/k_{0}=0.4\Delta$.
\end{description}

To our best knowledge, the superluminal Focused X wave has not realized
experimentally yet, but probably the approach worked out for luminal Focus
Wave Modes recently~\cite{KaidogaFWM} may help to accomplish that.

\section{Discussion and conclusions}

Let us make finally some remarks on the intriguing superluminality issues of
the X-type wave pulses. Indeed, while\ phase velocities greater than $c$ are
well known in various fields of physics, a superluminal group velocity more
often than not is considered as a taboo, because at first glance it seems to
be at variance with the special theory of relativity, particularly, with the
relativistic causality. However, since the beginning of the previous century
-- starting from Sommerfeld's works on plane-wave pulse propagation in
dispersive media and precursors appearing in this process -- it is known that
group velocity need not to be a physically profound quantity and by no means
should be confused with signal propagation velocity. But in case of X-type
waves not only the group velocity exceeds $c$ but the pulse as whole
propagates rigidly faster than $c$.

A diversity of interpretations concerning this startling but experimentally
verified fact~\cite{PRLmeie},\cite{ItaalExp} can be encountered. On the one
end of the scale are claims, based on a sophisticated mathematical
consideration, that the relativistic causality is violated in case of these
pulses~\cite{Brasiil}. A recent paper~\cite{ShaaCausal} devoted to this issue
proves, however, that the causality is not violated globally in the case of
the X-type waves, but still the author admits a possibility of noncausal
signalling locally.

On the opposite end of the scale are statements insisting that the pulse is
not a real one but simply an interference pattern rebuilt at every point of
its propagation axis from truly real plane wave constituents travelling at a
slight tilt with respect to the axis. Such argumentation is not wrong but
brings us nowhere. Of course, there is a similarity between superluminality of
the X wave and a faster-than-light movement of the cutting point in the
scissors effect or of a bright stripe on a screen when a plane wave light
pulse is falling at the angle $\theta$ onto the screen plane. But in the
central highest-energy part of the X wave there is nothing moving at the tilt
angle. The phase planes are perpendicular to the axis and the whole field
moves rigidly along the axis. The Pointing vector lays also along the axis,
however, the energy flux is not superluminal. Hence, to consider the X waves
as something inferior compared to ''real'' pulses is not sound. Similar logic
would bring one to a conclusion that femtosecond pulses emitted by a
mode-locked laser are not real but ''simply an interference'' between the
continuous-wave laser modes. In other words, one would ignore the
superposition principle of linear fields, which implies reversible relation
between ''resultant'' and ''constituent'' fields and does not make any of
possible orthogonal basis inferior than others. Moreover, even plane waves, as
far as they are truly real ones, suffer from a certain superluminality.
Indeed, as it is well known, the most simple physically feasible realization
of a plane wave beam is the Gaussian beam with its constrained cross-section
and, correspondingly, a finite energy flux. However, one can readily check on
the analytical expressions for the beam (see also Ref.~\cite{Gouy}) that due
to the Gouy phase shift the group velocity in the waist region of the Gaussian
beam is slightly superluminal.

We are convinced that the X-type waves are not -- and cannot be -- at variance
with the special theory of relativity since they are derived as solutions to
the D'Alambert wave equation and corresponding electromagnetic vector fields
are solutions to the Maxwell equations. The relativistic causality has been
inherently built into them as it was demonstrated also in the present paper,
when we developed the Sommerfeld representation basing upon the
relativistically invariant retarded and advanced Green functions. An analysis
of local evolution and propagation of a ''signal mark'' made, e. g. by a
shutter onto the X wave is not a simple task due to diffractive changes in the
field behind the ''mark''. Therefore conclusions concerning the local
causality may remain obscured. However,\ a rather straightforward geometrical
analysis in the case of infinitely wideband X wave (with the width parameter
$\Delta\rightarrow0$ ) shows that the wave cannot carry any causal signal
between two points along its propagation axis. So, we arrive at conclusion
that the X-type waves constitute one example of ''allowed'' but nontrivial
superluminal movements. As a matter of fact -- although perhaps it is not
widely known -- superluminal movements allowed by the relativistic causality
have been studied since the middle of the previous century (see references in
~\cite{GinzReview}). For example, the reflection of a light pulse on a
metallic planar surface could be treated as 2-dimensional Cherenkov-Mach
radiation of a supeluminal current induced on the surface. In the same vain,
the representation of the X waves as generated by the Huygens-type sources
might be developed further, vis. we could place a real wire along the
propagation axis and treat the outgoing cone of the wave as a result of
cylindrical reflection of (or of radiation by the superluminal current in the
wire induced by) the leading collapsing cone of the wave.

In conclusion, superluminal movement of individual material particles is not
allowed but excitations in an ensemble may propagate with any speed, however,
if the speed exceeds $c$ they cannot transmit any physical signal. Last two
decades have made it profoundly clear how promising and fruitful is studying
of the superluminal phenomena instead of considering them as a sort of
trivialities or taboos. We have in mind here not only the localized waves or
photon tunneling \ or propagation in inverted resonant media, etc., but also
-- or even first of all -- the implementation and application of entangled
states of Einstein-Podolsky-Rozen pairs of particles in quantum
telecommunication and computing.

This research was supported by the Estonian Science Foundation Grant No.3386.
The author is very grateful to the organizers of this exceptionally
interesting Conference in warm atmosphere of Naples.

\bigskip

\begin{thebibliography}{99}
\bibitem {slingshot}Mathematical procedures of derivation of wave fields as if
they are generated by sources from complex locations (i.e. a coordinate is
formally made a complex number) are well known in the theory of the Gaussian
beams already as well as in treating the localized waves. For example, in the
Ref.\cite{ZioMAIN} an expression for the zeroth-order X wave had been obtained
as Li\'{e}nard-Wiechert\ potentials of an electron moving along the complex
$z$ axis displaced from real space, but this comparatively formal approach
remained undeveloped.
\end{thebibliography}
\end{document}